\documentclass[lettersize,journal]{IEEEtran}
\usepackage{amsmath,amsfonts}
\usepackage{amssymb}
\usepackage{nicefrac}
\usepackage{algorithmic}
\usepackage{algorithm}
\usepackage{array}
\usepackage[caption=false,font=normalsize,labelfont=sf,textfont=sf]{subfig}
\usepackage{textcomp}
\usepackage{stfloats}
\usepackage{url}
\usepackage{xspace}
\usepackage{verbatim}
\usepackage{graphicx}
\usepackage{cite}
\usepackage{xcolor}
\newcommand{\Cv}{C_\mathrm{V}}
\newcommand{\adjustedaccent}[1]{%
	\mathchoice{}{}
	{\mbox{\raisebox{-.5ex}[0pt][0pt]{$\scriptstyle#1$}}}
	{\mbox{\raisebox{-.35ex}[0pt][0pt]{$\scriptscriptstyle#1$}}}
}
\newcommand{\bow}[1]{\overset{\adjustedaccent{\smallfrown}}{#1}}
\newcommand{\dd}{\mathrm{d}}
\newcommand{\qtd}{\mathrm{Q3D}}

\begin{document}
	
	\title{Quasi-3D Magneto-Thermal Quench Simulation Scheme for Superconducting Accelerator Magnets}
	
	\author{Laura A.~M.~D'Angelo, Yvonne Späck-Leigsnering, and Herbert De Gersem
		\thanks{L.~A.~M.~D'Angelo, Y.~Späck-Leigsnering, and H.~De Gersem are with the Institut für Teilchenbeschleunigung und Elektromagnetische Felder (TEMF) and with the Centre for Computational Engineering, Technische Universität Darmstadt, Darmstadt 64289, Germany (e-mail: dangelo@temf.tu-darmstadt.de).}%
	}
	
	\markboth{This work has been submitted to the IEEE for possible publication.}%
	{MT27-Q7SF84RL}
	
	
	\maketitle
	
	\begin{abstract}
		To tackle the strong multi-scale problem in the quench simulation of superconducting accelerator magnets, this work proposes a hybrid numerical method which uses two-dimensional first-order finite-elements in the magnet cross-section and one-dimensional higher-order orthogonal polynomials in longitudinal direction. 
	\end{abstract}
	
	
	\section{Introduction}
	\IEEEPARstart{S}{uperconducting} cables can suffer from the quench phenomenon which is a sudden shift to normal-conducting state~\cite{Wilson_1983aa}. 
	As a consequence, heat generating electric losses arise in the quenched volume, which can lead to a thermal runaway. In 2008, this phenomenon caused massive damages and a shutdown of the LHC at CERN~\cite{Bajko_2009aa}.
	
\begin{figure}[tbp]
		\centering 
		\includegraphics[width=1\columnwidth]{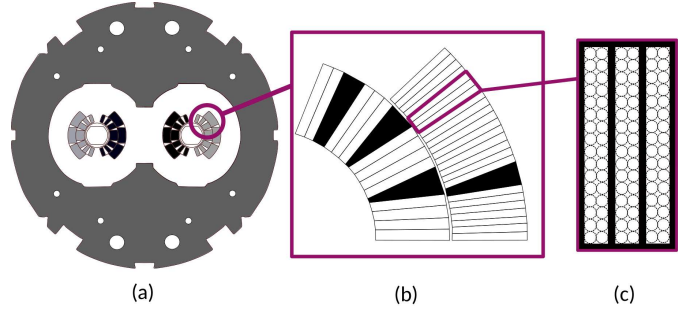}
		\caption{In (a), the cross-section of the LHC main dipole is shown. The coil configuration is depicted in (b) with its multi-filamentary Rutherford cables in (c). Figure adapted from~\cite{DAngelo_2020aa}.}
		\label{fig:dipole_geometry}
	\end{figure}

	Numerical field simulation is a powerful tool to investigate and predict the quenches in superconducting magnets.
	However, conventional numerical methods struggle with the geometry of accelerator magnets: The magnet cross-section contains many geometrical details (see Fig.~\ref{fig:dipole_geometry}a-c), which require a fine spatial resolution. 
	The length of the magnet is much larger than its diameter, e.g.~the length of LHC main dipole is twenty times its diameter, and thus it is a strong multi-scale problem. 
	Moreover, the quench phenomenon is a highly-nonlinear multi-physical problem~\cite{Russenschuck_2010aa}, for which fully coupled three-dimensional (3D) simulation is by far out of reach.
	Two-dimensional (2D) models, however, cannot cover the longitudinal quench propagation~\cite{Bortot_2016aa}.
	To remedy this, this paper develops a quasi-3D (Q3D) method for coupled magneto-thermal quench simulation, which has already been successfully applied for the thermal hot-spot propagation~\cite{DAngelo_2020aa} and
	magnetic field simulation~\cite{DAngelo_2021aa} in superconducting models. Thereby, we combine a conventional (2D) finite-element (FE) discretization in the transversal plane (see Fig.~\ref{fig:dipole_geometry}a-c) with a one-dimensional (1D) spectral-element (SE) discretization in the longitudinal direction. 
	First, superconducting magnets are described as a magneto-thermal problem, and the discretized Q3D model is derived. Finally, the approach is successfully verified against and compared to a conventional 3D FE simulation. 
	
	\section{Magneto-Thermal Formulation}
	The standard magnetoquasistatic (MQS) formulation reads
	\begin{equation}
	\nabla \times \left( \nu \nabla \times \vec{A} \right) = \vec{J}
	\label{eq:standard_mqs}
	\end{equation}
	with $\vec{A}$ being the magnetic vector potential (MVP) in Vs/m, $\nu$ the reluctivity in Vs/Am and $\vec{J}$ the current density in A/mm$^2$ which is decomposed as
	\begin{equation}
	\vec{J} = \vec{J}_\mathrm{s} + \vec{J}_\mathrm{Ohm} + \vec{J}_\mathrm{IFCC}.
	\label{eq:currents}
	\end{equation}
	Here, $\vec{J}_\mathrm{s}$ is the source current density, and $\vec{J}_\mathrm{Ohm}$ is the eddy current density occurring in normal-conducting regions and follows Ohm's rule, $\vec{J}_\mathrm{Ohm} = \sigma \vec{E}$, with $\sigma$ as electrical conductivity in S/m and $\vec{E}$ as electrical field strength in V/m~\cite{Jackson_1998aa}. Furthermore, the interfilament coupling current (IFCC) phenomena occurring in multifilamentary superconducting cables is taken into account by
	\begin{equation}
	\vec{J}_\mathrm{IFCC} = -\nu\tau \partial_t \vec{B},
	\end{equation}
	where $\tau$ is the characteristic IFCC time constant in s and $\vec{B}=\nabla \times \vec{A}$ is the magnetic field density in T~\cite{Verweij_1995aa}. Persistent currents and interstrand coupling currents~\cite{Verweij_1995aa} are not considered in this work, but can be included in the numerical scheme analogously to the IFCCs. By expressing $\vec{J}_\mathrm{Ohm}$ and $\vec{J}_\mathrm{IFCC}$ in terms of the MVP, the MQS formulation \eqref{eq:standard_mqs} becomes 
	\begin{equation}
	\nabla \times \left( \nu \nabla \times \vec{A} \right) + \sigma \partial_t \vec{A} + \nabla \times \left( \nu \tau \nabla \times \partial_t \vec{A} \right) = \vec{J}_\mathrm{s}.
	\label{eq:mqs}
	\end{equation}
	
	The heat conduction equation reads
	\begin{equation}
	-\nabla \cdot \left( \lambda \nabla \vartheta \right) + \Cv \partial_t \vartheta = q,
	\label{eq:heat_conduction}
	\end{equation}
	where $\vartheta$ is the temperature in K, $\lambda$ is the thermal conductivity in W/m/K, $\Cv$ is the volumetric heat capacity in J/m$^3$/K, and $q$ is the heat source density in W/m$^3$. The MQS formulation~\eqref{eq:mqs} is linked with~\eqref{eq:heat_conduction} through the heat losses arising from the currents. Hence, the heat source density is decomposed as
	\begin{equation}
	q = \underbrace{\rho |\vec{J}_s|^2}_{=q_\mathrm{s}} + \underbrace{ \sigma |\partial_t \vec{A}|^2 }_{=q_\mathrm{Ohm}} + \underbrace{ \nu\tau |\nabla \times \partial_t \vec{A}|^2}_{=q_\mathrm{IFCC}}
	\label{eq:heatlosses}
	\end{equation}
	with $\rho$ as electrical resistivity in Ohm m. All quantities in \eqref{eq:mqs} and \eqref{eq:heat_conduction} depend on the spatial coordinate $\vec{r} = (x,y,z)$ in m and the time $t$. Additionally, the materials are nonlinear in terms of the temperature $\vartheta$ as explained in detail in Sec.~\ref{sec:nonlinear}.
	
	\section{Quasi-3D Method}
	\subsection{Discretization}
	In the Q3D setting, the model's geometry is decomposed in a transversal $xy$-plane (the magnet's cross-section) and a longitudinal $z$-direction (along the magnet's length). Then, the cross-section is discretized with 2D triangular FEs, while 1D spectral line elements using orthogonal polynomials are employed for the longitudinal direction. The MVP is split up into a transversal component $\vec{A}_t$ lying in the $xy$-plane and a longitudinal component $\vec{A}_\ell$ pointing in $z$-direction, which are approximated by~\cite{DAngelo_2021aa}
	\begin{equation}
	\vec{A} = \begin{bmatrix}
	\vec{A}_t \\ \vec{A}_\ell 
	\end{bmatrix} \approx \sum\limits_{ejqw} 
	\begin{bmatrix}
	\bow{a}^{(t)}_{eq}(t) \; \vec{w}_e(x,y) \phi_q^k(z) \\
	\bow{a}^{(\ell)}_{jw}(t) \; N_j(x,y) \phi_w^k(z)
	\end{bmatrix}\; \forall k,
	\label{eq:q3d_mvp}
	\end{equation}
	and the temperature is discretized as~\cite{DAngelo_2020aa}
	\begin{equation}
	\vartheta \approx \sum\limits_{jq} u_{jq}(t) \; N_j(x,y) \phi_q^k(z) \quad \forall k.
	\label{eq:q3d_temp}
	\end{equation}
	Herein, $N_j$ and $\vec{w}_e$ are first order 2D FE nodal and edge functions on the $j$-th node and $e$-th edge~\cite{Brenner_2008aa}, respectively, and $\phi_q^k$ are orthogonal polynomials of $q$-th order on the $k$-th SE~\cite{Shen_2011aa}. This work chooses $\phi_q^k$ as the modified Lobatto polynomials introduced in~\cite{Fakhar-Izadi_2015aa} due to their beneficial properties for the Q3D method cf.~\cite{DAngelo_2021aa}. 
	By applying the Ritz-Galerkin approach in \eqref{eq:mqs}-\eqref{eq:heat_conduction}, and inserting the discretizations \eqref{eq:q3d_mvp}-\eqref{eq:q3d_temp}, one obtains the semi-discrete Q3D systems of equations
	\begin{align}
	\mathbf{K}^{\qtd,\mathrm{e}}_\nu \mathbf{a} + \left( \mathbf{K}^{\qtd,\mathrm{e}}_{\nu\tau} + \mathbf{M}^{\qtd,\mathrm{e}}_\sigma \right) \frac{\dd \mathbf{a}}{\dd t} &= \mathbf{j}^{\qtd,\mathrm{e}}_\mathrm{s}, \label{eq:mag_system} \\
	\mathbf{K}^{\qtd,\mathrm{n}}_\lambda \mathbf{u} + \mathbf{M}^{\qtd,\mathrm{n}}_{\Cv} \frac{\dd \mathbf{u}}{\dd t} &= \mathbf{q}^{\qtd,\mathrm{n}}, \label{eq:th_system}
	\end{align}
	with the discrete Q3D quantities
	\begin{align}
	{\mathbf{a}} &= \left[ {\mathbf{a}}^{(t)}, \: {\mathbf{a}}^{(\ell)} \right]^\mathsf{T}, \label{eq:mvp_solution} \\
	\mathbf{K}^\mathrm{Q3D,e}_\alpha &= 
	\begin{bmatrix}
	\mathbf{K}^\mathrm{Q3D,t}_\alpha & -\mathbf{D}^\mathrm{SE}_\alpha \otimes \mathbf{C}^\mathrm{FE,t\ell}_\alpha \\ 
	\left( -\mathbf{D}^\mathrm{SE}_\alpha \otimes \mathbf{C}^\mathrm{FE,t\ell}_\alpha \right)^\mathsf{T} & \mathbf{K}^\mathrm{Q3D,\ell}_\alpha  
	\end{bmatrix}, \label{eq:q3d_curlcurl} \\
	\mathbf{K}^\mathrm{Q3D,t}_\alpha &= \mathbf{M}^\mathrm{SE}_\alpha \otimes \mathbf{K}^\mathrm{FE,t}_\alpha + \mathbf{K}^\mathrm{SE}_\alpha \otimes \mathbf{M}^\mathrm{FE,t}_\alpha, \\
	\mathbf{K}^\mathrm{Q3D,\ell}_\alpha &= \mathbf{M}^\mathrm{SE}_\alpha \otimes \mathbf{K}^\mathrm{FE,\ell}_\alpha, \\
	\mathbf{M}^\mathrm{Q3D,e}_\sigma &= 
	\begin{bmatrix}
	\mathbf{M}^\mathrm{SE}_\sigma \otimes \mathbf{M}^\mathrm{FE,t}_\sigma & \mathbf{0} \\ 
	\mathbf{0} & \mathbf{M}^\mathrm{SE}_\sigma \otimes \mathbf{M}^\mathrm{FE,\ell}_\sigma 
	\end{bmatrix}, \\
	\mathbf{j}^\mathrm{Q3D,e}_\mathrm{s} &= \left[ \mathbf{0}, \: \mathbf{q}_1^\mathrm{SE} \otimes (\mathbf{X}^\mathrm{FE,\ell} \mathbf{i}_\mathrm{s}) \right]^\mathsf{T}, \\ 
	\mathbf{K}^\mathrm{Q3D,n}_\lambda &= \mathbf{M}^\mathrm{SE}_\lambda \otimes \mathbf{K}^\mathrm{FE,n}_\lambda + \mathbf{K}^\mathrm{SE}_\lambda \otimes \mathbf{M}^\mathrm{FE,n}_\lambda, \\
	\mathbf{M}^\mathrm{Q3D,n}_{C_\mathrm{V}} &= \mathbf{M}^\mathrm{SE}_{C_\mathrm{V}} \otimes \mathbf{M}^\mathrm{FE,n}_{C_\mathrm{V}}, \\
	\mathbf{q}^\mathrm{Q3D,n} &= \mathbf{q}^\mathrm{SE} \otimes \mathbf{q}^\mathrm{FE,n}, \label{eq:losses_q3d}
	\end{align}
	where the superscript $\mathrm{SE}$ indicates the 1D SE matrices, $(\mathrm{FE,t})$ the 2D FE transversal edge matrices, $(\mathrm{FE,\ell})$ the 2D FE longitudinal edge matrices, while $\mathbf{K}$ is a corresponding stiffness matrix, $\mathbf{M}$ a mass matrix, $\mathbf{D}$ a damping matrix, $\mathbf{C}$ an edge-coupling matrix~\cite{DAngelo_2021aa}, and $\mathbf{X}$ a winding matrix~\cite{Schops_2013aa}. The vectors $\mathbf{a}$ and $\mathbf{u}$ collect the Q3D DoFs of the MVP \eqref{eq:q3d_mvp} and temperature \eqref{eq:q3d_temp}, respectively. These matrices and vectors are discussed in detail in \cite{DAngelo_2020aa} and \cite{DAngelo_2021aa}.
	In \eqref{eq:q3d_curlcurl}-\eqref{eq:losses_q3d}, we observe:
	\begin{enumerate}
		\item The Kronecker tensor product $\otimes$~\cite{VanLoan_2000aa} between standard 1D SE and 2D FE entities allows for efficient assembly.
		\item The secondary diagonals in the Q3D curlcurl matrix \eqref{eq:q3d_curlcurl} are nonzero. Thus, the transversal and longitudinal MVP components in \eqref{eq:mvp_solution} are coupled with each other, i.e.~system \eqref{eq:mag_system} indeed describes a 3D physical behavior.
	\end{enumerate}
	Finally, the problem is discretized in time with the implicit Euler scheme. 
	
	\begin{figure}[tbp]
		\centering 
		\includegraphics[width=1\columnwidth]{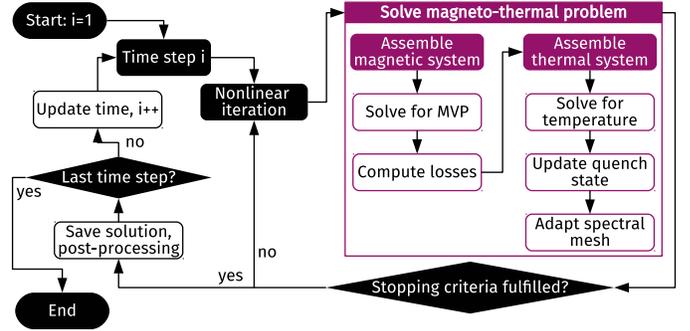}
		\caption{Magneto-thermal coupling scheme: In each time step a nonlinear iteration is performed in which the coupled problem (purple box) is solved at the current working point.}
		\label{fig:magneto-thermal-coupling-diagram}
	\end{figure}
	
	\subsection{Solving Procedure}
	The magneto-thermal coupled problem is solved numerically according to the purple marked scheme in Fig.~\ref{fig:magneto-thermal-coupling-diagram}: The magnetic system \eqref{eq:mag_system} is assembled at the current working point and solved for the MVP. In a post-processing step, the discrete heat losses based on \eqref{eq:heatlosses} are computed and used as right-hand side for the thermal system, for which the matrices are also assembled at the current working point. After the system is solved for the temperature, the quench state of the model is updated. This information is used to decide if an adaptation of the mesh in $z$-direction is necessary to ensure a good resolution of the steep temperature gradients. This procedure is embedded in an efficient nonlinear iteration scheme (see Sec.~\ref{sec:nonlinear}) and a time stepping loop (both visualized in black in Fig.~\ref{fig:magneto-thermal-coupling-diagram}).
	
	\section{Nonlinear Iteration Scheme}\label{sec:nonlinear}
\begin{figure}[tbp]
		\centering 
		\includegraphics[width=1\columnwidth]{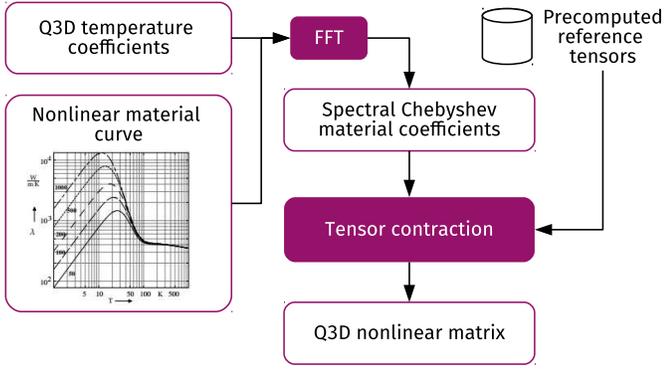}
		\caption{Scheme for an efficient treatment of nonlinearities in the Q3D setting: From nonlinear material curves and Q3D temperature coefficients, the Chebyshev material coefficients are obtained via FFT. Using precomputed reference tensors and a tensor contraction approach, the Q3D nonlinear matrix is efficiently assembled in each nonlinear iteration. Material curve from \cite{Russenschuck_2010aa}.}
		\label{fig:nonlinearity-diagram}
	\end{figure}

	The materials used in superconducting magnets -- copper, NbTi or Nb$_3$Tn, glass fibre, helium -- exhibit strong nonlinear behaviors in terms of the temperature~\cite{Russenschuck_2010aa}. Therefore, it is crucial to implement an efficient treatment of these nonlinearities for the Q3D method. To this end, a nonlinear material $\alpha$ is expanded in $z$-direction in terms of Chebyshev polynomials $T_m$ of $m$-th order~\cite{Shen_2011aa} instead of modified Lobatto polynomials,
	\begin{equation}
	\alpha(z) \approx \sum_m \widetilde{\alpha}_m \; T_m(z).
	\end{equation}
	Here, $\widetilde{\alpha}_m$ are the Chebyshev coefficients of $\alpha$ evaluated for a specific temperature distribution. The reason for preferring Chebyshev polynomials over modified Lobatto polynomials in this case is that the Chebyshev coefficients can be computed very efficiently, e.g.~by using a Fast Fourier Transform (FFT)~\cite{Shen_2011aa}. As a result, the 1D spectral matrices are calculated by solving an integral over a product of a Chebyshev polynomial and two modified Lobatto polynomials (or their derivatives), yielding the element-wise nonlinear 1D SE matrices 
	\begin{align}
	\left( \mathbf{K}^\mathrm{SE}_\alpha \right)^{(k)}_{pq} &= \sum\limits_m \widetilde{\alpha}_m \int\limits_{I_k}  T_m(z) \partial_z \phi^k_q(z) \partial_z \phi^k_p(z) \, \dd z, \label{eq:NL_stiffness} \\
	\left( \mathbf{M}^\mathrm{SE}_\alpha \right)^{(k)}_{pq} &= \sum\limits_m \widetilde{\alpha}_m \int\limits_{I_k}  T_m(z) \phi^k_q(z)\phi^k_p(z) \, \dd z, \\
	\left( \mathbf{D}^\mathrm{SE}_\alpha \right)^{(k)}_{pq} &= \sum\limits_m \widetilde{\alpha}_m \int\limits_{I_k}  T_m(z) \partial_z \phi^k_q(z) \phi^k_p(z) \, \dd z. \label{eq:NL_damping}
	\end{align}
	with $I_k$ being the subinterval of the $k$-th SE.
	For these triple polynomial integrals, closed forms do not exist. Hence, they have to be calculated using numerical integration. To avoid repeating these calculations in each nonlinear iteration, the integrals in \eqref{eq:NL_stiffness}-\eqref{eq:NL_damping} are precomputed for each combination $(p,q,m)$ and stored as 3D reference tensors in a database. In these precomputations, neither geometry nor material information is yet included. Then, the Q3D nonlinear matrices can be obtained by an efficient tensor contraction approach~\cite{Kirby_2007aa} using the Chebyshev material coefficients and adding the missing geometry data. This scheme is depicted in Fig.~\ref{fig:nonlinearity-diagram} and allows for an efficient assembly of the nonlinear Q3D matrices in each nonlinear iteration.
	
	\section{Verification of the Simulation Scheme}
	\subsection{Superconducting Multi-Filamentary Wire Model}
	\begin{figure}
		\centering 
		\includegraphics[width=.5\columnwidth]{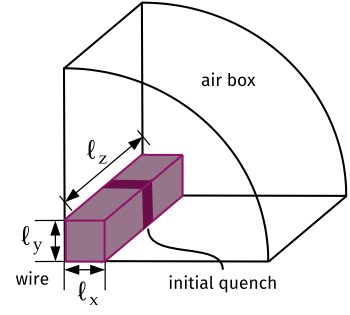}
		\caption{Computational model of the superconducting multi-filamentary wire used for verification with $\ell_x = \ell_y=10^{-4}\,$m and $\ell_z=1\,$m.}
		\label{fig:model}
	\end{figure}
	To verify the proposed nonlinear coupled scheme in the Q3D setting, a single quarter superconducting multi-filamentary wire with rectangular form $\ell_x \times \ell_y \times \ell_z$ is considered, where $\ell_x = \ell_y=10^{-4}\,$m and $\ell_z=1\,$m are the wire's width, height and length, respectively. An initial temperature distribution is defined such that a central segment along $z$ already quenches at the beginning of the simulation time window. The computational model is built as sketched in Fig.~\ref{fig:model}. For the magnetic simulation, an air box is put around the wire. For the thermal problem, it suffices to solve the heat conduction equation only in the wire domain, where isothermal temperatures are assumed at the wire fronts.
	
	\subsection{Quench State Model}
	\begin{figure}[tbp]
		\centering 
		\includegraphics[width=.8\columnwidth]{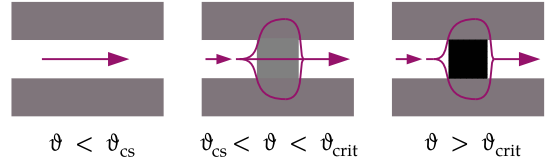}
		\caption{Current sharing phenomenon in a multi-filamentary superconducting cable. Superconducting regions are depicted in white, normal-conducting regions in shades of gray indicating their conductivity. Arrows visualize the transport current flow. }
		\label{fig:current-sharing}
	\end{figure}
\begin{figure}[tbp]
		\centering 
		\includegraphics[width=.7\columnwidth]{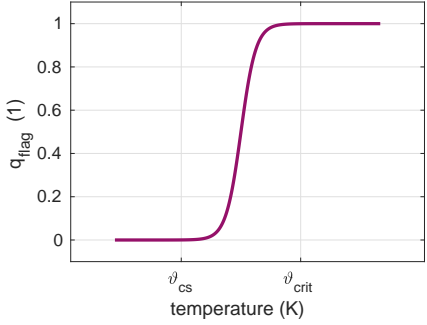}
		\caption{Smooth quench state model based on the current sharing and critical temperature of the superconductor.}
		\label{fig:qflag}
	\end{figure}

	The quench state is expressed by a temperature-dependent quench flag function $q_\mathrm{flag}(\vartheta) \mapsto [0,1]$ taking the current sharing phenomenon~\cite{Sharma_2021aa} into account which occurs in superconducting multi-filamentary wires, as visualized in Fig.~\ref{fig:current-sharing}: If the temperature in a superconducting filament rises above a current sharing temperature $\vartheta_\mathrm{cs}$, the filament starts to quench, i.e.~its resistivity rises. The transport current in this filament begins to favor paths in the surrounding copper matrix, although a part of the current still flows through the partly quenched filament segment. However, when the temperature surpasses the critical temperature $\vartheta_\mathrm{crit}$, the filament segment is shifted completely to normal-conducting state. As the electrical conductivity of NbTi and Nb$_3$Sn is lower than of copper, the transport current will completely flow around the quenched spot via the copper matrix. 
	This behavior is modeled with an appropriately constructed sigmoidal function,
	\begin{equation}
		q_\mathrm{flag}(\vartheta) = \left( 1 + \exp \left( -16 \frac{\vartheta - \vartheta_\mathrm{crit} }{ \vartheta_\mathrm{crit} - \vartheta_\mathrm{cs} } + 8 \right) \right)^{-1},
		\label{eq:qflag}
	\end{equation}
	which is plotted in Fig.~\ref{fig:qflag} and which we prefer over a simpler linear ramp function to avoid discontinuities which could disturb the convergence of the nonlinear iteration procedure. Then, the quench-state-dependent materials are given by
	\begin{align}
		\tau(\vartheta) &= (1 - q_\mathrm{flag}(\vartheta)) \, \tau_\mathrm{SC}, \\
		\sigma(\vartheta) &= q_\mathrm{flag}(\vartheta) \, \sigma_\mathrm{Cu}(\vartheta), \\
		\rho(\vartheta) &= q_\mathrm{flag}(\vartheta) \, \rho_\mathrm{Cu}(\vartheta),
	\end{align}
	with $\tau_\mathrm{SC}$ as IFCC time constant of the superconductor, $\sigma_\mathrm{Cu}$ and $\rho_\mathrm{Cu}$ as the electrical conductivity and resistivity of copper, respectively.
	
	\subsection{Comparison with GetDP}
\begin{figure}[tbp]
		\centering 
		\includegraphics[width=.9\columnwidth]{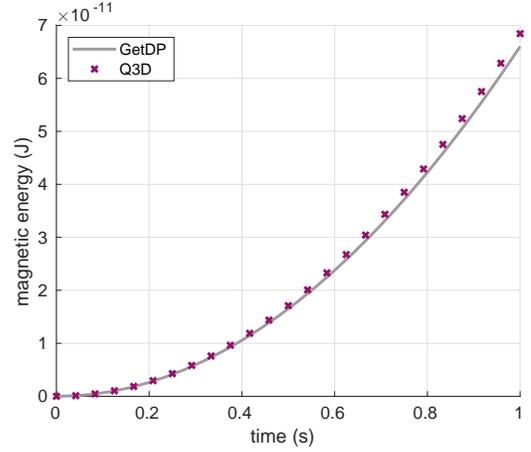}
		\caption{Magnetic energy over time computed by 3D FE GetDP and Q3D hybrid FE in-house solver.}
		\label{fig:magnetic_energy}
	\end{figure}

\begin{figure}[tbp]
		\centering 
		\includegraphics[width=.9\columnwidth]{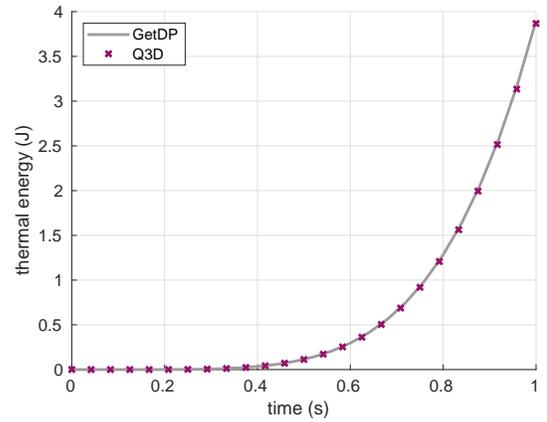}
		\caption{Thermal energy over time computed by 3D FE GetDP and Q3D hybrid FE in-house solver.}
		\label{fig:thermal_energy}
	\end{figure}

	\begin{table}[tbp]
		\centering 
		\caption{Comparison between 3D FE and Q3D hybrid method}
		\begin{tabular}{c|c|c}
			& \textbf{3D GetDP} & \textbf{Q3D solver} \\ \hline 
			\textbf{\#FE} & 1,863,431 & 330 \\ 
			\textbf{\#SE} & -- & 5 \\
			\textbf{Polynomial order} & -- & 6 \\
			\textbf{Magnetic \#DoF} & 1,741,669 & 21,979  \\ 
			\textbf{Thermal \#DoF} & 80,735 & 806 \\
			\textbf{Computation time} & 1\,h & 5\,min \\ \hline
		\end{tabular}
	\label{tab:comparison}
	\end{table}
	The superconducting multi-filamentary wire model is simulated with our Q3D in-house code and a conventional 3D FE method using the software GetDP~\cite{Dular_2019aa}. Attention has been paid to ensure that the numbers of FEs in the cross-section of both models are similar to allow for comparability. The time range of $[0,1]\,$s is run through a total of 25 constant time steps. The magnetic and thermal energy are computed for each time step in both solvers and compared to each other in Fig.~\ref{fig:magnetic_energy} and Fig.~\ref{fig:thermal_energy} showing an excellent agreement between the data. Tab.~\ref{tab:comparison} lists the number of SEs and the chosen polynomial order for the Q3D method, and compares the required number of FEs and degrees of freedom (DoF) as well as the computation time of both methods. Due to the strong multi-scale nature of the model, it is not surprising that the 3D model requires a high number of FEs to achieve an accurate meshing especially along the $z$-direction of the computational domain. On the other hand, the Q3D method employs higher-order SEs in the longitudinal direction which achieve a high accuracy with only a few elements. Hence, the Q3D method ends up with only approx.~$1\,\%$ as many DoFs as the conventional 3D FE method. Consequently, the Q3D simulation takes $5\,$min and the 3D simulation $1\,$h to run on a standard workstation.
	
	\section{Conclusion}
	A Q3D method with linear FEs in the transversal cross-section and higher-order SEs in the longitudinal direction has been proposed to tackle the multi-physical, multi-scale problem of quench simulation in superconducting accelerator magnets. Furthermore, an efficient scheme to deal with nonlinear material curves in the Q3D setting has been introduced. The method has been applied to the quench simulation of a superconducting multi-filamentary wire and has been successfully verified against a conventional 3D FE simulation. The results are promising regarding a possible application of the method to the quench simulation of superconducting magnets.

	\section*{Acknowledgments}
	This work has been supported by the German BMBF project "Quenchsimulation für Supraleitende Magnete: Steigerung der Auflösung in Zeit und Raum" (BMBF-05P18RDRB1), by the DFG Research Training Group 2128 "Accelerator Science and Technology for Energy Recovery Linacs", and by the Graduate School Computational Engineering at TU Darmstadt.

	\vfill
	
\end{document}